\documentclass[twocolumn]{aastex701}
\usepackage{amsmath} % For mathematical equations
\usepackage{amssymb} % For mathematical symbols
\usepackage{bm}      % For bold vectors
\usepackage{mathtools} % Extended math tools

\usepackage{xcolor}
\begin{document}

\title{Numerical Investigation of Efficient Electron Acceleration at an Unsteady Solar Flare Loop-Top}

\author[orcid=0000-0003-3735-944X,gname=Yoshiaki,sname=Sato]{Yoshiaki Sato}
\affiliation{Department of Astronomical Science, The Graduate University for Advanced Studies, SOKENDAI, 2-21-1, Osawa, Mitaka, Tokyo 181-8588, Japan}
\affiliation{National Astronomical Observatory of Japan, 2-21-1 Osawa, Mitaka, Tokyo 181-8588, Japan}
\email{yoshiaki.sato@grad.nao.ac.jp}

\author[orcid=0000-0002-7800-9262,gname=Takafumi,sname=Kaneko]{Takafumi Kaneko}
\affiliation{Faculty of Education, Niigata University, 8050 Ikarashi 2-no-cho, Nishi-ku, Niigata, 950-2181, Japan}
\email{kaneko@ed.niigata-u.ac.jp}

\author[orcid=0000-0002-6330-3944,gname=Noriyuki,sname=Narukage]{Noriyuki Narukage}
\affiliation{National Astronomical Observatory of Japan, 2-21-1 Osawa, Mitaka, Tokyo 181-8588, Japan}
\affiliation{Department of Astronomical Science, The Graduate University for Advanced Studies, SOKENDAI, 2-21-1, Osawa, Mitaka, Tokyo 181-8588, Japan}
\email{noriyuki.narukage@nao.ac.jp}

\author[orcid=0000-0003-3882-3945,gname=Shinsuke,sname=Takasao]{Shinsuke Takasao}
\affiliation{Humanities and Sciences/Museum Careers, Musashino Art University, Tokyo 187-8505, Japan}
\email{stakasao@musabi.ac.jp}

\begin{abstract}
Using magnetohydrodynamic (MHD) fields and guiding-center test-particle calculations, this study investigates how time-dependent loop-top dynamics modulate the adiabatic energization of electrons in a solar flare.
Our results indicate that a time-varying loop-top structure enhances acceleration efficiency compared to a quasi-steady one. In the quasi-steady velocity field, the net acceleration is suppressed due to the decelerating effect of betatron cooling. Conversely, in the unsteady velocity field, the betatron mechanism readily accelerates electrons within the compressed magnetic field at the edge of the loop-top.
These findings suggest that the acceleration of electrons at the loop-top is driven not only by the static shape of the magnetic structure but also by dynamic events such as plasmoid collisions. While previous studies have primarily focused on acceleration processes within the reconnection outflow, such as at termination shocks or within plasmoids, our research highlights the importance of the acceleration and deceleration processes at the exit point where electrons escape from the loop-top.
\end{abstract}


\section{Introduction} \label{sec:intro}
Solar flares are the most energetic phenomena in the solar system, releasing as much as $10^{32}$~erg of energy via magnetic reconnection on timescales ranging from minutes to hours \citep[for reviews, see e.g.,][]{benz_flare_2017}.
A significant fraction of this energy is converted into the kinetic energy of non-thermal particles, whose interactions with the ambient plasma environment generate radio, hard X-ray (HXR), and gamma-ray emissions \citep{aschwanden_particle_2002,fletcher_solar_2024}.
Understanding the timing, locations, mechanisms, and efficiency of this particle acceleration remains a key challenge in solar physics.

Early observations with the \textit{Yohkoh} satellite revealed HXR sources located above bright soft X-ray flare loops \citep[e.g.,][]{Masuda_1994}.
This feature, often termed the loop-top or above-the-loop-top (ALT) source and hereafter referred to simply as the "loop-top", appears to be a common characteristic of solar flares \citep{petrosian_loop_2002}.
Accelerated electrons were also observed around the diffusion region \citep{narukage_evidence_2014}.
These discoveries prompted investigations into particle acceleration mechanisms within the framework of the standard flare model \citep{shibata_hot-plasma_1995,priest_magnetic_2002}.
For a comprehensive review of particle acceleration processes in solar flares, see \citet{zharkova_recent_2011}.
Key mechanisms proposed to occur within the reconnection region and its outflow include acceleration at termination shocks \citep{aurass_shock-excited_2002,chen_particle_2015,kong_acceleration_2019}, stochastic acceleration in turbulence \citep{petrosian_stochastic_2012}, and energization within contracting magnetic islands (plasmoids) \citep{drake_electron_2006,oka_electron_2010,zank_particle_2014}.

Loop-top acceleration has also been discussed in the framework of collapsing magnetic traps.
In this picture, reconnected field lines form a magnetic trap that can energize particles near the loop-top \citep{Somov_1997}.
Kinematic models have used time-dependent electromagnetic fields to follow particle orbits and energization in collapsing traps, from the original framework to later 2.5D and 3D extensions \citep{Giuliani_2005,Grady_2009,Grady_2012}.
Related studies have addressed loss-cone evolution, particle escape, relativistic energization, and acceleration in collapsing traps with braking plasma jets \citep{EradatOskoui_2014a,EradatOskoui_2014b,Borissov_2016}.
Recent work has also examined the relative roles of betatron and Fermi energization in collapsing magnetic traps \citep{Mowbray_2025}.
Together, these studies show that particle energization is closely linked to magnetic-trap evolution and provide the context for our use of two MHD loop-top states to examine how loop-top flows modify the local adiabatic energy-gain terms.

Observations further show that the loop-top is not a static structure but a dynamic environment, continually perturbed by reconnection outflows and the embedded plasmoids they carry.
The motion and collision of plasmoids correlate temporally with impulsive radio and HXR bursts \citep{nishizuka_particle_2015,takasao_observational_2016}, and the arrival of reconnection outflows at the loop-top is associated with bursts of HXR and microwave emissions \citep{asai_downflow_2004,chen_particle_2015,yu_magnetic_2020}.

Numerical models further support this picture of a dynamic loop-top.
Magnetohydrodynamic (MHD) simulations first showed that reconnection outflows, even without plasmoids, establish oscillating, shock-filled structures at the loop-top \citep{takasao_magnetohydrodynamic_2015,takasao_above-the-loop-top_2016}.
This dynamic picture, including the presence of termination shocks, turbulent interfaces, and oscillations, has been further investigated and confirmed in subsequent numerical studies \citep[e.g.,][]{cai_investigations_2019,xie_numerical_2022,shen_origin_2022,Shibata_2023}.
Additionally, other simulations demonstrate that plasmoid collisions amplify turbulence at the loop-top and create highly dynamic conditions \citep{jelinek_oscillations_2017,ye_role_2020}.
Within this turbulent environment, particles confined in magnetic trapping structures can be efficiently accelerated through repeated shock interactions \citep{kong_acceleration_2019}.
Recent observations have confirmed that loop-top regions are intrinsically unsteady, exhibiting phenomena such as periodic pulsations \citep{nakariakov_quasi-periodic_2006,French_2024}, non-thermal velocities indicative of turbulence \citep{Ashfield_2024}, anisotropic turbulent flows driven by MHD instabilities \citep{xie_2025}, and electron acceleration within magnetic bottle structures \citep{chen_energetic_2024}.
Reconnection outflows further perturb the loop-top region, driving plasma flows and oscillations that coexist with non-thermal particle populations \citep{reeves_hot_2020,Shibata_2023}.
Collectively, these findings establish unsteady dynamics as a ubiquitous characteristic of flare loop-tops.

Several recent studies have combined MHD simulations with test-particle approaches to investigate electron acceleration in dynamically evolving reconnection and loop-top regions \citep[e.g.,][]{kong_dynamical_2020,kong_numerical_2022,bacchini_particle_2024,chen_energetic_2024}.
Additionally, comprehensive three-dimensional (3D) MHD simulations have explored multidimensional effects such as loop-top turbulence formation via the Kelvin-Helmholtz instability \citep{ruan_magnetohydrodynamic_2023}, turbulent interface regions below termination shocks \citep{shen_origin_2022}, and chromospheric responses \citep{druett_exploring_2024}.
These studies have advanced our understanding of particle acceleration, trapping, and transport in dynamic magnetic structures.
However, they primarily focus on acceleration mechanisms within reconnection regions, particle confinement within loop-top structures, or transport through the overall loop system.
A detailed quantitative assessment of how the transient evolution of the loop-top specifically modulates the energy gain of electrons as they escape from the loop-top, particularly at the exit point where betatron acceleration and Fermi reflection processes can either counteract or reinforce each other, remains to be explored.
To address this gap, we contrast the net energy gain of electrons in a quasi-steady loop-top against that in an unsteady loop-top perturbed by a plasmoid collision.
Through a combination of 2.5-dimensional (2.5D) MHD simulation and test-particle calculations, we decompose particle energy gains into contributions from Fermi reflection and betatron processes.
Throughout this paper, we use "Fermi reflection" as a descriptive label for the deterministic, adiabatic parallel energization associated with repeated reflections between converging magnetic mirrors, following related discussions of reflection-driven Fermi energization in reconnection and contracting magnetic structures \citep[e.g.,][]{Birn_2004, drake_electron_2006, oka_electron_2010, zank_particle_2014, arnold_electron_2021, oka_particle_2023}. This usage distinguishes the process analyzed here from the stochastic acceleration associated with \citet{fermi_origin_1949}; the present calculation does not address power-law formation.
Our analysis reveals that the temporal evolution of the loop-top environment critically modulates particle acceleration efficiency.

\section{Method} \label{sec:method}
We employed a two-step numerical approach to investigate electron acceleration mechanisms at the loop-top. First, we performed a 2.5-dimensional MHD simulation of a flare driven by magnetic reconnection. In the second step, we conducted a test-particle simulation using the magnetic and electric fields obtained from the MHD simulation as background fields. This method allowed us to track in detail the trajectories and energy changes of electrons within the plasma structures generated by the MHD simulation.

We numerically solved the MHD equations including nonlinear heat conduction along magnetic field lines as described in \citet{2017ApJ...845...12K}.

\subsection{MHD simulation} \label{subsec:simulation}
We defined the simulation domain in a Cartesian coordinate system with $0<x<60~\mathrm{Mm}$ and $-50~\mathrm{Mm}<y<300~\mathrm{Mm}$, where $x$ and $y$ represent the directions parallel and perpendicular to the solar surface, respectively. The grid size was set to $75~\mathrm{km}$ in the $x$-direction and $150~\mathrm{km}$ in the $y$-direction.
All variables were assumed to be constant in the $z$-direction.
The vector variables have a $z$-component as well as $x$- and $y$-components.

We assumed symmetry about the $x=0$ axis and solved the equations only in the $x>0$ region.
At $x=0$, we imposed symmetry boundary conditions for $B_x$, $B_z$, $\rho$, $p$, and $v_y$, and anti-symmetry boundary conditions for $B_y$, $v_x$, and $v_z$.
At the right boundary ($x=60~\mathrm{Mm}$), the same symmetry or anti-symmetry conditions were applied to ensure that the reconnection flow remains confined within the domain.
A fixed boundary condition was applied to the bottom ($y=-50~\mathrm{Mm}$) and top ($y=300~\mathrm{Mm}$) boundaries.

Radiative cooling was not included, as the characteristic timescale of the simulation ($\sim 100~\mathrm{s}$) is shorter than the radiative cooling timescale ($\sim 10^3-10^4~\mathrm{s}$) for a flare loop with a density of $n \sim 10^{10}-10^{11}~\mathrm{cm^{-3}}$ and a temperature of $T \sim 10^7~\mathrm{K}$.
The gravity $\boldsymbol{g}$ was given as
\begin{equation}
  \boldsymbol{g}=-\frac{GM_{\odot}}{\left(R_{\odot} + y\right)^{2}}\boldsymbol{e}_{y},
\end{equation}
where $\boldsymbol{e}_{y}$ is the unit vector in the $y$-direction,
$G$ is the gravitational constant, $M_{\odot}$ is the solar mass and $R_{\odot}$ is the solar radius.

The initial temperature profile was given as
\begin{equation}
  T(y)=T_{\mathrm{p}}+\frac{T_{\mathrm{c}}-T_{\mathrm{p}}}{2}
  \left[ 1+\tanh \left( \frac{y+2w_{0}}{w_{0}}\right)\right],
\end{equation}
where $T$ represents the temperature, $T_{\mathrm{p}}=10^{4}~\mathrm{K}$,
$T_{\mathrm{c}}=2\times 10^{6}~\mathrm{K}$, and $w_{0}=3~\mathrm{Mm}$.
Note that $y=0$ corresponds to the base of the corona.
The initial gas pressure and density profiles were calculated assuming hydrostatic equilibrium as
\begin{equation}
    p(y)=p_{\mathrm{c}}\exp \left[ -\int _{0} ^{y} \frac{mg(y^{\prime})}{k_{B}T(y^{\prime})}dy^{\prime}\right],
\end{equation}
\begin{equation}
    \rho (y)=\frac{mp(y)}{k_{B}T(y)}.
\end{equation}
where $p$ and $\rho $ represent the gas pressure and the mass density, respectively,
$m=8.3\times 10^{-25}~\mathrm{g}$ is the mean molecular mass for fully ionized hydrogen (half the proton mass),
and $k_{B}$ is the Boltzmann constant. The gas pressure at $y=0$ was given as $p_{c}=k_{B}n_{c}T_{c}$, where $n_{c}=2\times 10^{9}~\mathrm{cm^{-3}}$.
The initial magnetic field was given as
\begin{align}
  B_{x} &= 0,\\
  B_{y} &= -B_{0}\tanh (x/w_{1}), \\
  B_{z} &= B_{0}/\cosh (x/w_{1}),
\end{align}
where $B_{0}=20~\mathrm{G}$ and $w_{1}=3~\mathrm{Mm}$.
This force-free configuration keeps $B_y^2+B_z^2 = B_0^2$, so the magnetic pressure is spatially uniform and the guide-field component is concentrated around the initial current sheet.
The plasma beta is $\beta \equiv 8\pi p_{c}/B_{0}^{2} \sim 0.04$.

To initiate magnetic reconnection, we introduced a localized resistivity until $t=10~\mathrm{s}$, given as
\begin{equation}
  \eta = \eta _{0} \exp \left[ -\frac{x^{2}+(y-y_{0})^{2}}{\delta ^{2}}\right],
\end{equation}
where $\eta _{0}=5\times 10^{14}~\mathrm{cm^{2}~s^{-1}}$, $y_{0}=60~\mathrm{Mm}$ and $\delta = 0.5~\mathrm{Mm}$. After $t=10~\mathrm{s}$, the resistivity was switched to a uniform value
$\eta =10^{12}~\mathrm{cm^{2}~s^{-1}}$.
The Lundquist number $L_q$ is estimated to be $L_q = L_s v_A / \eta \sim 1.0 - 4.0 \times 10^{6}$, where $L_s$ is the density scale height $L_s = 120~\mathrm{Mm}$ and $v_A$ is the Alfv\'en velocity. Here, we take into account that the Alfv\'en velocity varies from $v_A = 1500 - 3000~\mathrm{km~s^{-1}}$ from the bottom of the corona to the upper corona due to the gravitational stratification.

We solved the MHD and heat conduction parts separately using the operator splitting method \citep{1968SJNA....5..506S}. The MHD part was solved using the four-step Runge-Kutta method \citep{2005A&A...429..335V,2017AIAAJ..55.1487J} and a fourth-order central finite difference scheme with an artificial viscosity \citep{2014ApJ...789..132R}. The heat conduction part was solved using the super time-stepping method \citep{2012MNRAS.422.2102M,2014JCoPh.257..594M} and a second-order central finite difference method with a slope limiter for anisotropic conductivity \citep{2007JCoPh.227..123S}.
To reduce the numerical errors caused by non-zero $\nabla \cdot \bm{B}$, we applied the hyperbolic divergence cleaning method \citep{2002JCoPh.175..645D}.

\subsection{Test-Particle Simulations}
We performed test-particle simulations using the results from the MHD simulation obtained in Section \ref{subsec:simulation} as the background fields. Under the typical physical conditions of the solar corona, the Alfv\'en timescale $\tau_A$ ($\sim 1 \mathrm{\ s}$) is significantly longer than the electron gyroperiod $\omega_{ce}^{-1}$ ($\sim 10^{-9}$ s), and the MHD grid spacing $\Delta x_{\mathrm{MHD}}$ ($\sim 10^4$ m) is significantly larger than the electron gyroradius $r_{ce}$ ($\sim 10^{-2}$ m). Therefore, the application of the guiding-center approximation (GCA) is justified for calculating particle trajectories.
We note that our test-particle GCA approach does not include microscopic physics, such as high-frequency waves and kinetic instabilities \citep[e.g.,][]{che_electron_2020}, as well as particle-particle interactions and feedback on the MHD fields. This approach is therefore not a fully kinetic model of flare electron acceleration. Although this means we neglect these micro-scale processes, preceding studies have demonstrated that the GCA remains a useful approach even in complex environments for quantifying the adiabatic component of particle acceleration \citep[e.g.,][]{Dahlin_2014, gordovskyy_combining_2019}. We consider that even in the turbulent fields of solar flares, adiabatic acceleration (Fermi reflection and betatron) associated with MHD-scale structural changes still has a contribution, and the GCA effectively isolates and quantifies this adiabatic component \citep[e.g.,][]{bacchini_particle_2024, oyre_test_2025}, which is the primary focus of this study.
To track electron energization, we numerically integrate the GCA system of equations, which includes relativistic effects \citep{Northrop_1963,Ripperda_2018}:
\begin{gather}
\begin{split}
  \frac{d \boldsymbol{r}}{d t} &= u_{\|} \boldsymbol{b}+\boldsymbol{u}_{E}+\frac{m_{e} c \gamma}{q \kappa^{2} B} \boldsymbol{b} \times\left(u_{\|} \frac{d \boldsymbol{b}}{d t}+\frac{d \boldsymbol{u}_{E}}{d t}\right) \\
  &\quad +\frac{\mu c}{q \gamma \kappa^{2} B} \boldsymbol{b} \times \nabla(\kappa B),
  \end{split} \\
  \frac{d\left(\gamma u_{\|}\right)}{d t}=\frac{q}{m_{e}} \boldsymbol{E}^{*} \cdot \boldsymbol{b}+\gamma \boldsymbol{u}_{E} \cdot \frac{d \boldsymbol{b}}{d t}-\frac{\mu}{m_{e} \gamma} \boldsymbol{b} \cdot \nabla(\kappa B) \label{eq: eom}
\end{gather}
where $\boldsymbol{u}_{E}$ is the $\boldsymbol{E} \times \boldsymbol{B}$ drift velocity, and $\frac{d \boldsymbol{b}}{d t}$ and $\frac{d \boldsymbol{u}_{E}}{d t}$ are the time derivatives of the magnetic field unit vector and the drift velocity, respectively, defined as:
\begin{gather}
  \boldsymbol{u}_{E}=c \frac{\boldsymbol{E}^{*} \times \boldsymbol{B}}{B^{2}}, \\
  \frac{d \boldsymbol{b}}{d t}=u_{\|}(\boldsymbol{b} \cdot \nabla) \boldsymbol{b}+\left(\boldsymbol{u}_{E} \cdot \nabla\right) \boldsymbol{b}, \\
  \frac{d \boldsymbol{u}_{E}}{d t}=u_{\|}(\boldsymbol{b} \cdot \nabla) \boldsymbol{u}_{E}+\left(\boldsymbol{u}_{E} \cdot \nabla\right) \boldsymbol{u}_{E}
\end{gather}
Furthermore, the dimensionless parameters describing relativistic effects are:
\begin{gather}
  \kappa=\sqrt{1-\left(\frac{u_{E}}{c}\right)^{2}}, \\
  \mu=\frac{m_{e}\left(\gamma u_{\perp}\right)^{2}}{2 B}, \label{eq: mu} \\
  \gamma=\frac{1}{\sqrt{1-\left(u_{\|}^{2}+u_{\perp}^{2}\right) / c^{2}}}
\end{gather}
Here, $\boldsymbol{r}$ is the guiding-center position vector, $u_{\|}$ is the electron velocity parallel to the magnetic field, $u_{\perp}$ is the gyration velocity, and $\boldsymbol{u}_{E}$ is the $\boldsymbol{E} \times \boldsymbol{B}$ drift velocity.
$c$ is the speed of light, $m_{e}$ is the electron mass, $q=-e$ is the electron charge, $B=|\boldsymbol{B}|$ is the magnetic field strength, and $\mu$ is the magnetic moment, which is assumed to be constant. To prevent unphysical electron acceleration parallel to the magnetic field due to anomalous resistivity, we define the electric field $\boldsymbol{E}^{*}$ by excluding the resistive term, following \citet{Birn_2017}, as:
\begin{equation}
\boldsymbol{E}^{*}=-\frac{1}{c} \boldsymbol{v} \times \boldsymbol{B} .
\end{equation}
We integrate the above equations using a four-step Runge-Kutta method \citep{2005A&A...429..335V,2017AIAAJ..55.1487J}. A uniform time step $\Delta t = \Delta x/c$ is adopted, where $\Delta x = 75~\mathrm{km}$ is the grid size in the $x$-direction of the MHD simulation. For each run of the test-particle simulations, the background MHD fields are assumed to be static. Thus, the particle orbits are calculated in frozen MHD snapshots rather than in fully time-dependent electromagnetic fields. At each particle time step, the field values at the guiding-center position are obtained by linear interpolation of the MHD grid cell values.

The initial conditions for the test particles are shown in Table \ref{tab:Conditions-for-test-particle-simulation}. The initial velocity distribution is thermal (Maxwellian), determined by the local temperature and density from the MHD simulation. The MHD simulation results were sampled every $2\,\mathrm{s}$ over the period $t = 100$--$200\,\mathrm{s}$.
\begin{table}[b]
  \caption{Initial conditions and parameters for test-particle simulations}
  \label{tab:Conditions-for-test-particle-simulation}
  \centering
  \scriptsize
  \setlength{\tabcolsep}{1pt}
  \begin{tabular}{@{}l@{\hspace{1.2em}}l@{}}
    \hline
    Parameter & Condition \\
    \hline
    Total number of particles & \parbox[t]{0.43\columnwidth}{\raggedright $10^7$} \\
    Velocity distribution & \parbox[t]{0.43\columnwidth}{\raggedright Maxwellian (determined by the $T$ and $\rho$ of the MHD background)} \\
    Spatial distribution & \parbox[t]{0.43\columnwidth}{\raggedright Proportional to $\rho$ in the current sheet and loop-top regions} \\
    Position & \parbox[t]{0.43\columnwidth}{\raggedright Current sheet + Loop-top} \\
    Pitch-angle ($\theta$) & \parbox[t]{0.43\columnwidth}{\raggedright Uniformly random $(0^\circ \leq \theta \leq 180^\circ)$} \\
    Tracking time ($t_{\mathrm{GCA}}$) & \parbox[t]{0.43\columnwidth}{\raggedright $1.0 \mathrm{\ s} (\leq \tau_A)$} \\
    \hline
  \end{tabular}
\end{table}
To evaluate the adiabatic electron energization at the loop-top, we analyzed the time evolution of the kinetic energy based on its physical origins. In GCA, the energy gain of a particle is described as the work done by the induced electric field $\boldsymbol{E}^*$ on the guiding center of the particle \citep{Northrop_1963}. Specifically, the adiabatic energy change in a magnetic trap can be decomposed into terms associated with Fermi reflection and betatron acceleration, which correspond to changes in kinetic energy parallel and perpendicular to the magnetic field, respectively \citep{Birn_2017}. Fermi reflection arises when particles bounce between approaching magnetic mirrors (contracting magnetic loops or plasmoids). As the trap contracts, particles gain parallel kinetic energy through repeated reflections under macroscopic converging flows \citep[e.g.,][]{Birn_2004, drake_electron_2006, oka_electron_2010, arnold_electron_2021, oka_particle_2023}. Within the GCA framework, the corresponding parallel energization is represented by the curvature-related term $m_{e} u_{\|}^2 \boldsymbol{u}_{E} \cdot ((\boldsymbol{b} \cdot \nabla) \boldsymbol{b})$ \citep[e.g.,][]{Northrop_1963, Birn_2012, Dahlin_2014, Birn_2017, li_acceleration_2021, zhou_electron_2015, oka_particle_2023}. Betatron acceleration, in contrast, arises from grad-B drift driven by magnetic field gradients; perpendicular energy increases as particles move into regions of stronger magnetic field or as the field strength at the particle's location increases over time.

To quantify the energy change, we adopt the non-relativistic limit ($\gamma \approx 1$, $\kappa \approx 1$).
We confirmed that most electrons remain non-relativistic (energy $\ll m_e c^2 \approx 511$ keV) throughout the tracking period.
Combining Eq.~\ref{eq: eom} and Eq.~\ref{eq: mu} yields the time evolution of $E_\parallel$ and $E_\perp$ as follows:
\begin{gather}
\frac{d E_{\|}}{d t}=m_{e} u_{\|}^2 \boldsymbol{u}_{E} \cdot ((\boldsymbol{b} \cdot \nabla) \boldsymbol{b}) -\mu u_{\|} \boldsymbol{b} \cdot \nabla B, \label{eq: dEpara} \\
\frac{d E_{\perp}}{d t}=\frac{d(\mu B)}{d t}=\mu u_{\|}(\boldsymbol{b} \cdot \nabla) B+\mu\left(\boldsymbol{u}_{E} \cdot \nabla\right) B. \label{eq: dEperp}
\end{gather}
The time evolution of the total kinetic energy $E=E_{\parallel}+E_{\perp}$ is the sum of these two, and is expressed as:
\begin{equation}
\frac{d E}{d t}=m_{e} u_{\|}^2 \boldsymbol{u}_{E} \cdot ((\boldsymbol{b} \cdot \nabla) \boldsymbol{b})+\mu\left(\boldsymbol{u}_{E} \cdot \nabla\right) B \label{eq: kinetic_energy},
\end{equation}
where the first term on the right-hand side of Eq.~\ref{eq: kinetic_energy} corresponds to Fermi reflection, and the second term corresponds to betatron acceleration \citep{Birn_2017}. The net energy gain from each acceleration mechanism is calculated by time-integrating each term over the tracking time $t_{\mathrm{GCA}}$:
\begin{gather}
\Delta E_{\text {fermi}}\left(t_{\mathrm{GCA}}\right)=\int_{0}^{t_{\mathrm{GCA}}} m_{e} u_{\|}^2 \boldsymbol{u}_{E} \cdot ((\boldsymbol{b} \cdot \nabla) \boldsymbol{b}) dt, \\
\Delta E_{\text {betatron}}\left(t_{\mathrm{GCA}}\right)=\int_{0}^{t_{\mathrm{GCA}}} \mu\left(\boldsymbol{u}_{E} \cdot \nabla\right) B d t.
\end{gather}

\section{Results} \label{sec:results}

\subsection{Overview}

We present an overview of the MHD simulation results described in Section~\ref{subsec:simulation}. The MHD data were sampled every $2\,\mathrm{s}$ over the period $t = 100$--$200\,\mathrm{s}$ for use in the test-particle calculations.

\begin{figure}[t]
  \centering
  \includegraphics[width=\columnwidth]{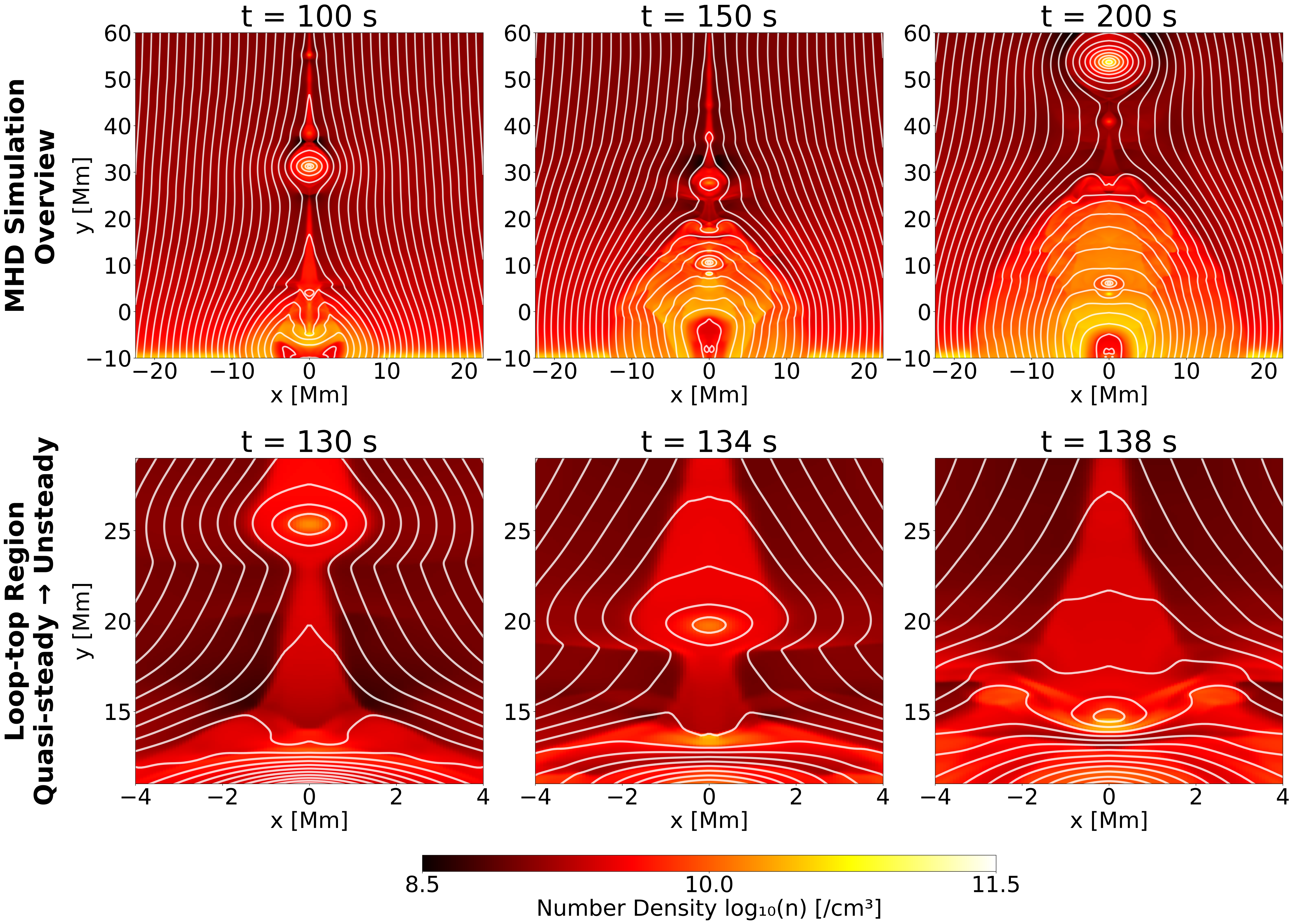}
  \caption{Time evolution of the number density distribution in the reconnection region from the MHD simulation. The upper panels show the overall view at $t = 100$, $150$, and $200\,\mathrm{s}$, illustrating the large-scale evolution of the reconnection region. The lower panels show zoomed-in views of the loop-top region at $t = 130$, $134$, and $138\,\mathrm{s}$, highlighting the transition from a quasi-steady state ($t = 130\,\mathrm{s}$) to an unsteady state ($t = 138\,\mathrm{s}$) caused by a plasmoid collision. White contours represent magnetic field lines.}
  \label{fig:mhd}
\end{figure}

Figure~\ref{fig:mhd} displays the time evolution of the number density distribution in the reconnection region. The upper panels illustrate the large-scale evolution of the reconnection region at $t = 100$, $150$, and $200\,\mathrm{s}$. The lower panels zoom in on the loop-top region at $t = 130$, $134$, and $138\,\mathrm{s}$, highlighting the transition from a quasi-steady configuration to an unsteady configuration driven by a plasmoid collision.

As reconnection proceeds, a downward outflow collides with the flare loop, forming a loop-top structure. The geometry of this loop-top evolves over time, and this evolution substantially influences the efficiency of electron acceleration. We extract two characteristic phases from this temporal evolution for detailed analysis. The quasi-steady state ($t = 130\,\mathrm{s}$) is characterized by a stable apparent geometry of the loop-top. The unsteady state ($t = 138\,\mathrm{s}$) occurs when a plasmoid, generated in the reconnection region, collides with the loop-top, causing a substantial deformation of the loop-top geometry, as shown in the lower panels of Figure~\ref{fig:mhd}. We analyze the behavior of electrons in each state in detail.

A comparative analysis of electron trajectories and their energy gain mechanisms in the quasi-steady and unsteady states was performed to elucidate the role of plasmoid collisions in electron acceleration.
The trajectories shown below are selected examples used to illustrate the acceleration mechanisms; their selection criteria are detailed in Appendix \ref{sec:appendix_criteria}.
Appendix \ref{sec:appendix_population} further presents a population analysis in a loop-top-edge region defined from the MHD fields, using a calculation with loop-top-only initial conditions.
We note that the electrons analyzed here are tracked from the termination of the outflow, and thus have not undergone the Fermi reflection processes within the outflow itself. In contrast, \citet{Dahlin_2014} examined electrons that experienced such Fermi reflection within the outflow.

\subsection{Analysis of the Quasi-steady State}\label{subsec:quasi-steady}

\begin{figure*}[!htbp]
\centering
\includegraphics[width=0.8\textwidth,height=0.85\textheight,keepaspectratio]{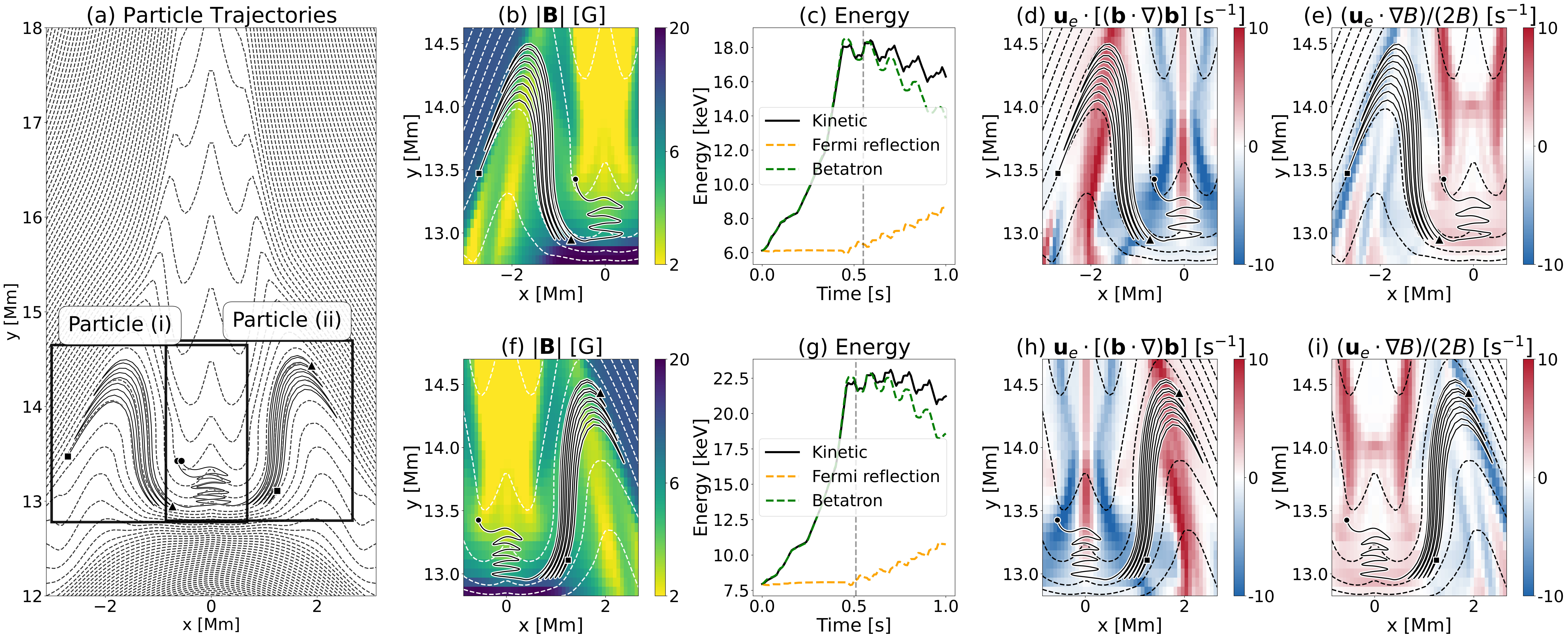}
\caption{
Electron acceleration in the quasi-steady loop-top configuration at $t = 130 \mathrm{\ s}$.
This figure shows the trajectories and energy evolution for two selected electrons.
(a) Overview of electron trajectories with magnetic field lines.
(b, f) Trajectories on the magnetic-field strength ($B$) map.
(c, g) Time evolution of kinetic energy (black), cumulative Fermi reflection energy gain (orange), and cumulative betatron energy gain (green).
(d, h) Trajectories overlaid on the rate of energy gain from Fermi reflection, $\boldsymbol{u}_E \cdot [(\boldsymbol{b} \cdot \nabla)\boldsymbol{b}]$.
(e, i) Trajectories overlaid on the betatron acceleration rate, $(\boldsymbol{u}_E \cdot \nabla B) / (2B)$.
In all panels, squares ($\blacksquare$) and circles ($\bullet$) denote the initial ($t=0$ s) and final ($t=1$ s) positions, respectively. Triangles ($\blacktriangle$) in the spatial maps (b, d, e, f, h, i) and dashed lines in the time plots (c, g) mark the onset of particle trapping at the loop-top edge.
In this state, betatron cooling (negative contribution) largely counteracts Fermi reflection, resulting in inefficient net energy gain.
}
\label{fig:steady_combined}
\end{figure*}

We analyzed the behavior of electrons in the quasi-steady state at $t=130\,\mathrm{s}$, when the loop-top structure is stable.
Unless otherwise specified, all $(x, y)$ coordinates are given in units of Mm.
In this state, electrons are trapped within the loop-top region, as shown by the particle trajectories in Figure~\ref{fig:steady_combined}, panel (a).
This trapping is facilitated by a magnetic mirror configuration, which is visible in the magnetic field maps in Figure~\ref{fig:steady_combined}, panels (b) and (f).
The loop-top region is characterized by this trap structure, which exhibits a weaker magnetic field along its central axis (e.g., extending from $(-2.2, 13.0)$ to $(-1.5, 14.5)$) and a stronger magnetic field at the trap boundaries (e.g., from $(-2.8, 13.5)$ to $(-2.0, 14.5)$, and near $(-1.0, 13.0)$), as visible in Figure~\ref{fig:steady_combined}, panel (b).
The energy of these trapped electrons exhibits periodic oscillations, and their net energy change shows either a decrease or stagnation.

The energy decomposition (Figure~\ref{fig:steady_combined}, panels (c) and (g)) reveals that the Fermi reflection component (orange curves) provides a net energy gain, whereas the betatron component (green curves) results in a net energy loss (e.g., from $t=0.5~\mathrm{s}$ to $1.0~\mathrm{s}$ in panel (g)).
For particles (i) and (ii), the initial energy gain (before the dashed line) reflects acceleration within the reconnection outflow; after trapping at the loop-top edge (marked by the triangle in panel (a)), the mechanisms analyzed here dominate, leading to energy stagnation.

\begin{figure*}[!htbp]
\centering
\includegraphics[width=0.8\textwidth,height=0.85\textheight,keepaspectratio]{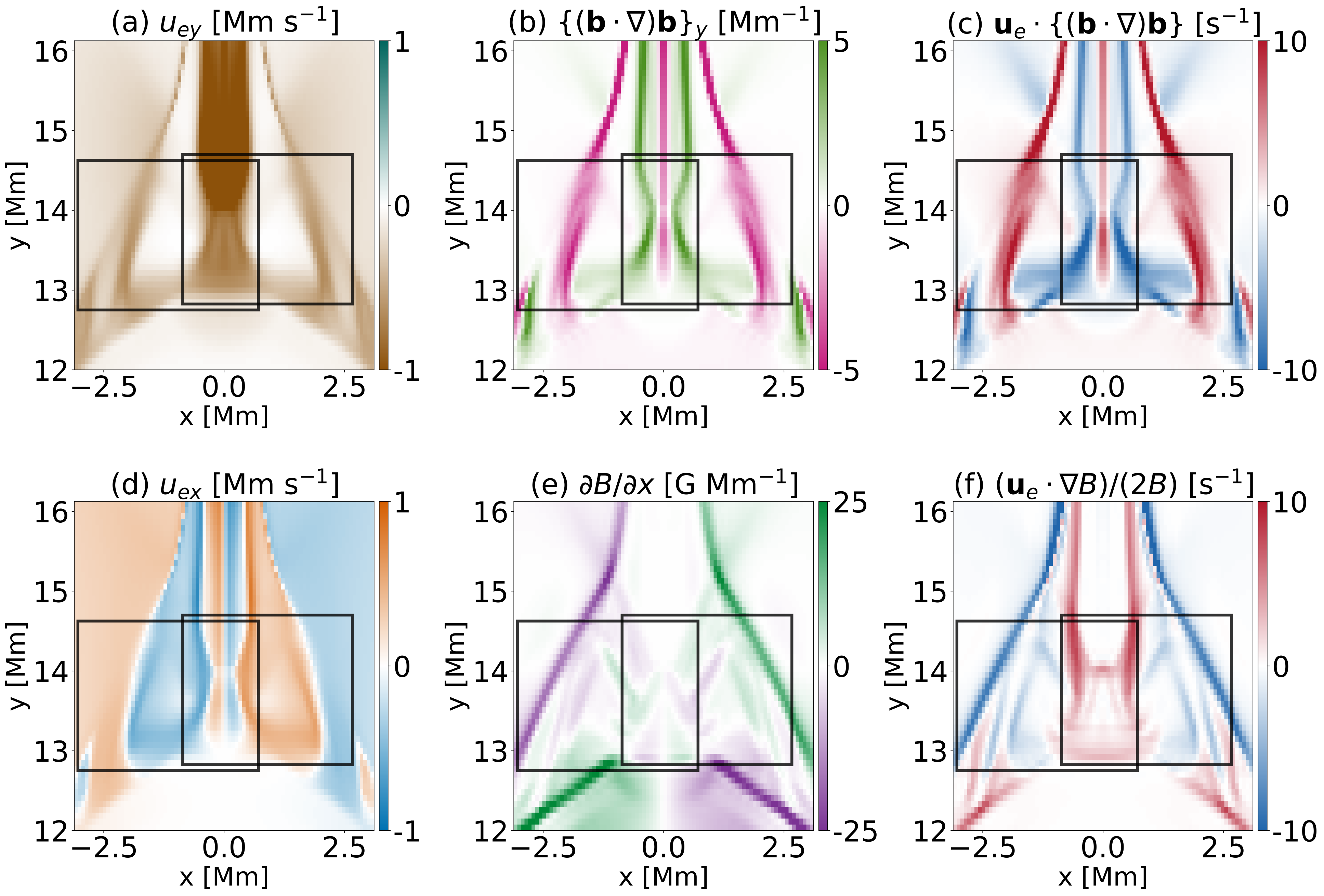}
\caption{Spatial distributions of quantities governing acceleration mechanisms at $t=130~\mathrm{s}$.
Panels (a)-(c) show factors for Fermi reflection: (a) vertical plasma flow ($u_{ey}$), (b) magnetic curvature ($[(\boldsymbol{b}\cdot\nabla)\boldsymbol{b}]_y$), and (c) the resulting acceleration rate.
Panels (d)-(f) show factors for betatron acceleration: (d) transverse plasma flow ($u_{ex}$), (e) magnetic field gradient ($\partial B/\partial x$), and (f) the resulting acceleration rate.
Black boxes mark the selected electron trajectory regions from Figure~\ref{fig:steady_combined}.}
\label{fig:detail_background_steady}
\end{figure*}

We investigated the spatial distribution of the acceleration rates determined by the background fields (Figure~\ref{fig:detail_background_steady}), where black boxes mark the trajectory regions of particles (i) and (ii) shown in Figure~\ref{fig:steady_combined}.
The rate of energy gain from Fermi reflection was predominantly positive (red regions) along the electron's trajectory (e.g., in the elongated region from $(-2.1, 13.8)$ to $(-1.4, 14.5)$ in Figure~\ref{fig:steady_combined}, panel (d)).
Conversely, the betatron term was largely negative (blue regions) along the trajectory (e.g., in the elongated region from $(-2.8, 13.0)$ to $(-2.0, 14.5)$ in Figure~\ref{fig:steady_combined}, panel (e)).
The physical origins of these opposing contributions are examined by analyzing the underlying plasma dynamics.

The positive rate of energy gain from Fermi reflection results from the alignment of the downward plasma flow with the downward curvature of the magnetic field lines.
In the quasi-steady state, a downward $\boldsymbol{E} \times \boldsymbol{B}$ drift exists ($u_{ey}<0$, see the brown area, e.g., in the elongated region from $(-2.1, 13.8)$ to $(-1.4, 14.5)$ in Figure~\ref{fig:detail_background_steady}, panel (a)), which corresponds to a downward plasma flow.
The loop-top region has a downward magnetic field line curvature ($[(\boldsymbol{b}\cdot\nabla)\boldsymbol{b}]_y<0$, see the pink area, e.g., in the elongated region from $(-2.1, 13.8)$ to $(-1.4, 14.5)$ in Figure~\ref{fig:detail_background_steady}, panel (b)).
The alignment of the downward plasma flow with the downward magnetic field curvature results in a positive rate of energy gain from Fermi reflection (red regions, e.g., in the region from $(-2.1, 13.8)$ to $(-1.4, 14.5)$ in Figure~\ref{fig:detail_background_steady}, panel (c)).

Betatron cooling originates from the configuration of the plasma's velocity field and the magnetic field gradient.
In the quasi-steady state, a slow shock forms downstream of the reconnection site. This shock drives a dominant inward plasma flow toward the loop-top region (e.g., $x \lesssim -2$, in Figure~\ref{fig:detail_background_steady}, panel (d)).
The loop-top is characterized by a magnetic field gradient that increases outward (e.g., $[\nabla B]_x<0$ in the region from $(-2.8, 13.5)$ to $(-2.0, 14.5)$ in Figure~\ref{fig:detail_background_steady}, panel (e)).
The inward flow in regions with a strong outward magnetic field gradient causes betatron cooling (blue regions, e.g., in the region from $(-2.8, 13.5)$ to $(-2.0, 14.5)$ in Figure~\ref{fig:detail_background_steady}, panel (f)) along the particle trajectories.

\begin{figure*}[!htbp]
  \centering
  \includegraphics[width=0.8\textwidth,height=0.85\textheight,keepaspectratio]{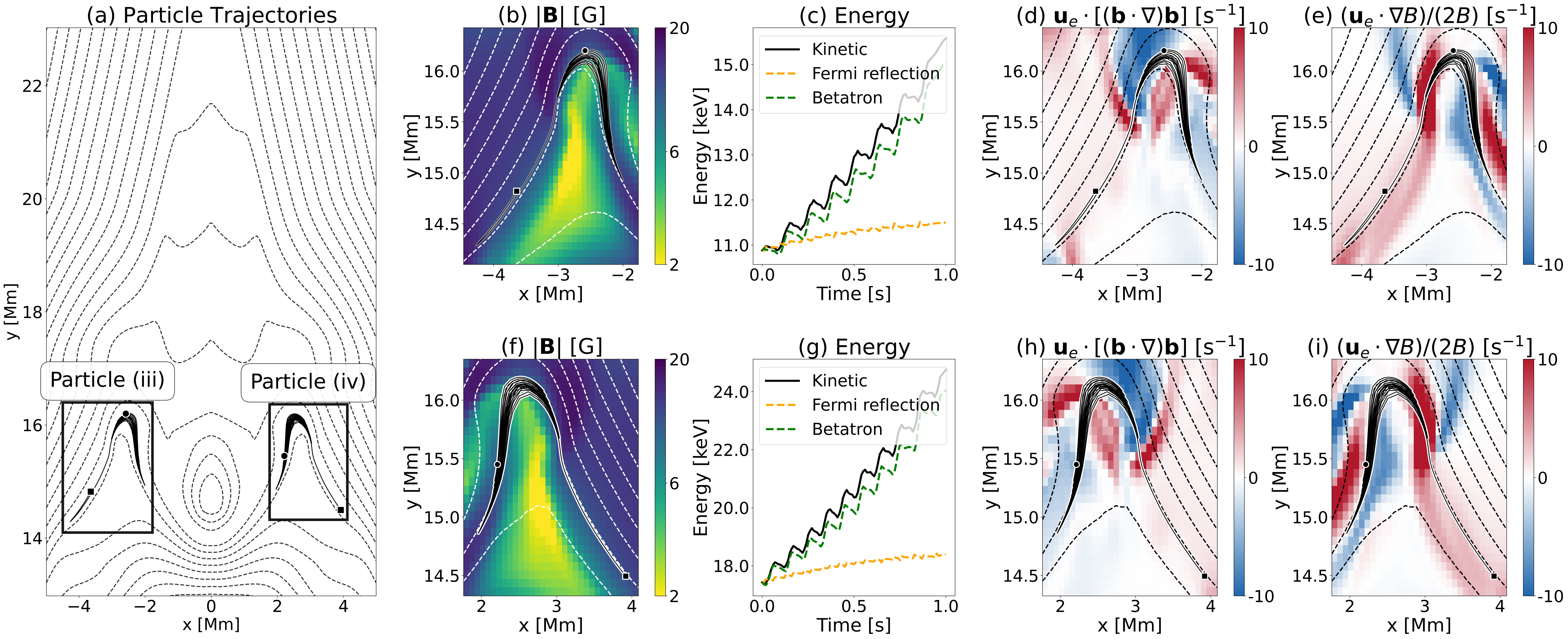}
  \caption{
Electron acceleration in the unsteady loop-top at $t = 138 \mathrm{\ s}$, perturbed by a plasmoid collision.
The panel layout is the same as in Figure~\ref{fig:steady_combined}.
In sharp contrast to the quasi-steady case, both Fermi reflection (orange) and betatron (green) gains are positive.
This synergy drives efficient net acceleration, shown by the rapid increase in total kinetic energy (black).
}
  \label{fig:unsteady_combined}
\end{figure*}

\begin{figure*}[!htbp]
  \centering
  \includegraphics[width=0.8\textwidth,height=0.85\textheight,keepaspectratio]{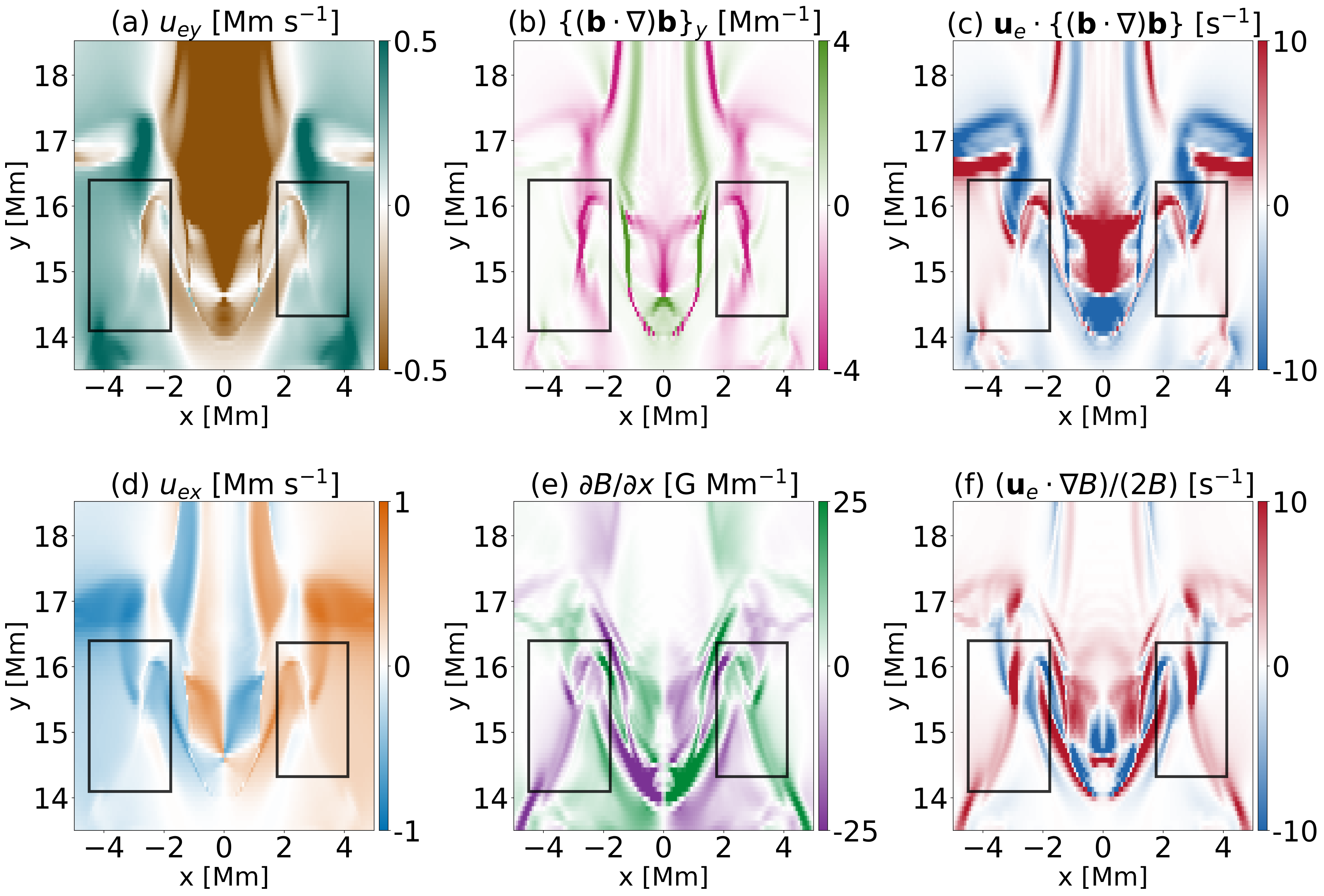}
  \caption{
Same as Figure~\ref{fig:detail_background_steady}, but for the unsteady loop-top at $t=138~\mathrm{s}$, perturbed by a plasmoid collision.
In contrast to the quasi-steady case, the outward plasma flow (d) now aligns with the magnetic field gradient (e), leading to positive betatron acceleration (f).
Black boxes mark the selected electron trajectory regions from Figure~\ref{fig:unsteady_combined}.
    }
  \label{fig:detail_background_unsteady}
\end{figure*}
\subsection{Analysis of the Unsteady State}\label{subsec:unsteady}

We analyze the behavior of electrons in the unsteady state at $t=138\,\mathrm{s}$, when a plasmoid collision has perturbed the loop-top (see the transition from the quasi-steady state at $t=130\,\mathrm{s}$ in Figure~\ref{fig:mhd}, lower panels).
Unless otherwise specified, all $(x, y)$ coordinates are given in units of Mm.

Similar to the quasi-steady state, electrons are trapped within the loop-top region, as shown by the particle trajectories in Figure~\ref{fig:unsteady_combined}, panel (a).
This trapping is facilitated by a magnetic mirror configuration, which is visible in the magnetic field maps in Figure~\ref{fig:unsteady_combined}, panels (b) and (f).
The loop-top region is characterized by this trap structure, which exhibits a weaker magnetic field along its central axis (e.g., from $(-3.1, 14.8)$ to $(-2.5, 16.0)$) and a stronger magnetic field at the boundaries (e.g., near $(-4.2, 14.2)$ and $(-2.0, 15.0)$), as visible in Figure~\ref{fig:unsteady_combined}, panel (b).
The temporal evolution of electron energy (Figure~\ref{fig:unsteady_combined}, panels (c) and (g)) contrasts sharply with that in the quasi-steady state. The electron energy (black curves) continuously increases, modulated by periodic oscillations.

The energy decomposition (Figure~\ref{fig:unsteady_combined}, panels (c) and (g)) reveals that, unlike in the quasi-steady state, both the Fermi reflection component (orange curves) and the betatron component (green curves) provide net positive contributions to the overall energy gain.

We investigated the spatial distribution of the acceleration rates determined by the background fields (Figure~\ref{fig:detail_background_unsteady}), where black boxes mark the trajectory regions of particles (iii) and (iv) shown in Figure~\ref{fig:unsteady_combined}.
The rate of energy gain from Fermi reflection (Figure~\ref{fig:unsteady_combined}, panels (d) and (h)) and the betatron acceleration rate (Figure~\ref{fig:unsteady_combined}, panels (e) and (i)) both show complex spatial structures, with electron trajectories repeatedly passing through both positive (red) and negative (blue) regions.
The physical origins of these alternating contributions are examined by analyzing the underlying plasma dynamics.

The complex pattern of the rate of energy gain from Fermi reflection arises from the interplay between plasma flow and magnetic curvature.
The vertical plasma flow $u_{ey}$ is not uniformly downward; it is upward ($u_{ey} > 0$, the green area) in the outward part of the trap (e.g., $x \lesssim -2.9$) and downward ($u_{ey} < 0$, the brown area) in the inner part (e.g., $x \gtrsim -2.9$), as shown in Figure~\ref{fig:detail_background_unsteady}, panel (a).
The loop-top region maintains a downward magnetic field line curvature ($[(\boldsymbol{b}\cdot\nabla)\boldsymbol{b}]_y < 0$, see the pink areas, e.g., in the region from $(-2.8, 14.5)$ to $(-2.5, 16.5)$ in Figure~\ref{fig:detail_background_unsteady}, panel (b)).
The combination of this bidirectional flow structure with the downward curvature results in acceleration (positive, red) in the inner region and deceleration (negative, blue) in the outer region, as seen in Figure~\ref{fig:detail_background_unsteady}, panel (c). The acceleration predominates, contributing to the net energy gain, because the region of strong downward curvature is skewed toward the inner (downward flow, $u_{ey} < 0$) region.

The betatron acceleration rate also shows a complex, oscillating pattern. The plasmoid collision induces bidirectional horizontal plasma flows ($u_{ex}$) within the trap, as shown in Figure~\ref{fig:detail_background_unsteady}, panel (d).
In the left black box region (Figure~\ref{fig:detail_background_unsteady}, panel (d)), the flow is entirely outward ($u_{ex} < 0$, the blue area), and is particularly strong (darker blue) in the rectangular region (e.g., $x \in [-3.5, -2.9]$ and $y \gtrsim 15.4$).
This flow combines with the magnetic field gradient, which is negative ($\partial B/\partial x < 0$, see the dark purple areas, e.g., in the region where $x \in [-3.5, -3.0]$ and $y \approx 15.5$ in Figure~\ref{fig:detail_background_unsteady}, panel (e)).
Consequently, the co-location of this strong outward flow (negative $u_{ex}$) and the strong negative magnetic gradient (negative $\partial B/\partial x$) creates a dominant region of intense positive acceleration ($\propto u_{ex}(\partial B/\partial x) > 0$), visible as the dark red area centered near $(-3.0, 15.5)$ in Figure~\ref{fig:detail_background_unsteady}, panel (f).
While regions of deceleration also exist (blue areas in Figure~\ref{fig:detail_background_unsteady}, panel (f)) where the gradient is positive ($\partial B/\partial x > 0$, see the green areas, e.g., in the region from $(-2.5, 15.5)$ to $(-2.0, 14.5)$ in Figure~\ref{fig:detail_background_unsteady}, panel (e)), the strong acceleration predominates, contributing to the net energy gain.

\subsection{Comparison of Quasi-steady and Unsteady States}

The preceding analysis of selected trajectories shows that the net change in electron energy differs between the two loop-top states. In the selected quasi-steady trajectories, the electron energy either stagnates or decreases. By contrast, in the selected unsteady trajectory, the electron energy increases continuously. These contrasting outcomes are caused by a sign reversal in the betatron contribution. The betatron mechanism switches from deceleration in the quasi-steady case to acceleration in the unsteady case.

Table~\ref{tab:acceleration_comparison} quantifies the energy change rates per oscillation cycle for selected particles in both states.
Individual oscillation cycles were identified by detecting local maxima in the temporal evolution of the electron's $x$-coordinate.
For each cycle, we calculated normalized energy change rates: the net kinetic energy change rate was normalized by the mean kinetic energy within that cycle, the Fermi reflection contribution by the mean parallel energy, and the betatron contribution by the mean perpendicular energy.
In the table, the first row for each state shows the mean value across all oscillation cycles, while the values in parentheses indicate the minimum and maximum values observed.

\begin{deluxetable*}{lcccc}[htbp]
  \tablecaption{Comparison of normalized energy change rates per oscillation cycle between quasi-steady and unsteady loop-top states
  \label{tab:acceleration_comparison}}
  \tablehead{
      \colhead{State} & \colhead{Particle} & \multicolumn{3}{c}{Energy change rate per cycle (\%)} \\
      \cline{3-5}
      & \colhead{ID} & \colhead{Net kinetic} & \colhead{Fermi reflection} & \colhead{Betatron}
  }
  \startdata
      Quasi-steady & (i) & $-2.7$ & $+11.7$ & $-8.9$ \\
      ($t=130\,\mathrm{s}$) & & $(-4.2 / -0.1)$ & $(+9.2 / +13.9)$ & $(-11.7 / -4.2)$ \\
      \hline
      Unsteady & (iii) & $+4.4$ & $+3.5$ & $+4.8$ \\
      ($t=138\,\mathrm{s}$) & & $(+4.0 / +4.9)$ & $(+1.3 / +6.8)$ & $(+3.8 / +6.0)$ \\
  \enddata
  \tablecomments{Energy change rates represent cycle-averaged values normalized by the corresponding energy component.}
\end{deluxetable*}

The main difference between the two cases analyzed here is the sign reversal of the betatron energy change rate. In the quasi-steady state, the betatron mechanism exhibits a negative mean rate (mean $-8.9\%$), acting as a cooling process. In the unsteady state, the betatron mechanism shows a positive mean rate (mean $+4.8\%$), contributing to acceleration. This sign reversal alters the net energy change: the quasi-steady case yields a negative net rate (mean $-2.7\%$), while the unsteady case achieves a positive net rate (mean $+4.4\%$). The betatron energy change rate is governed by the plasma flow velocity ($\boldsymbol{u}_{E}$). Thus, the direction and magnitude of the backflow of the reconnection outflow in the loop-top region control the sign of the betatron term.

\begin{figure*}[htbp]
  \centering
  \includegraphics[width=0.85\textwidth,height=0.75\textheight,keepaspectratio]{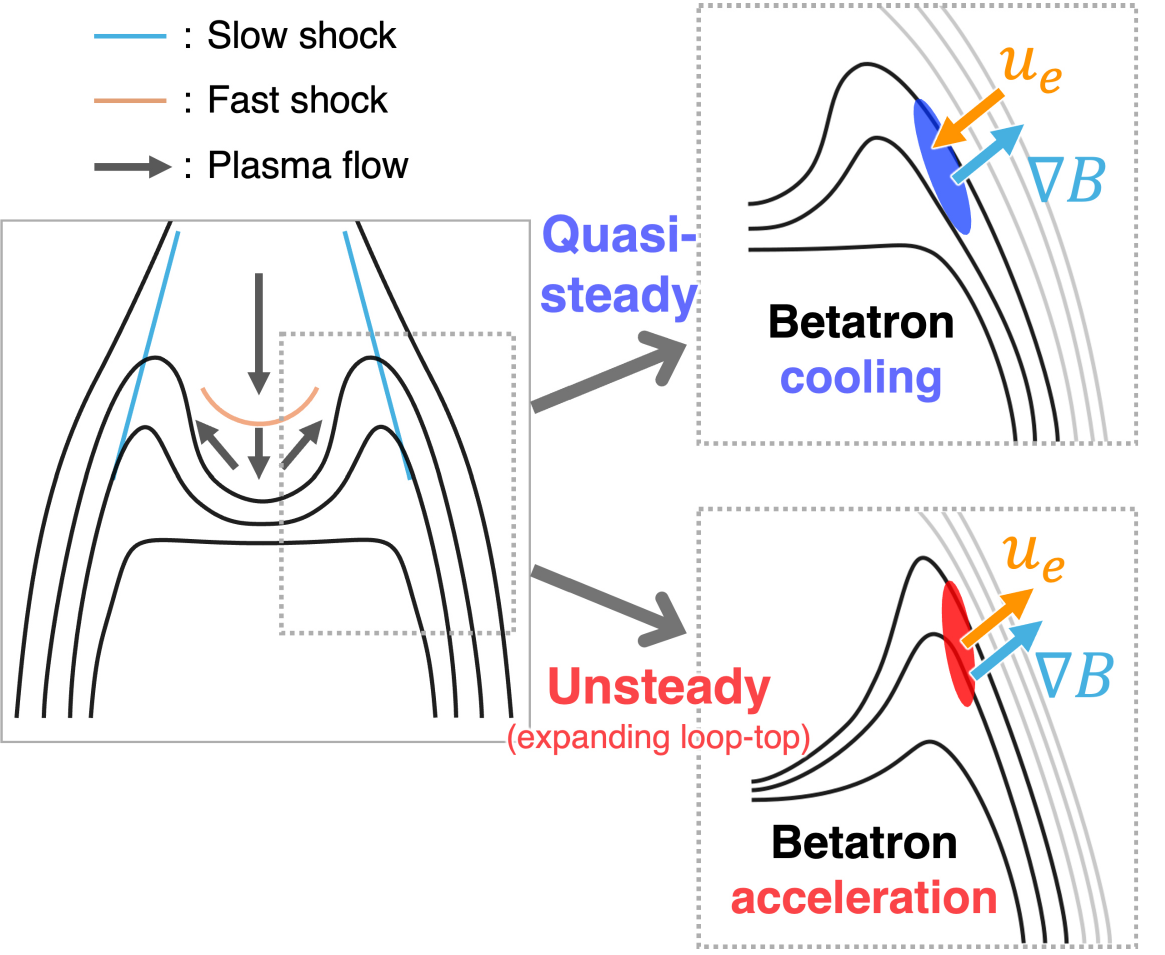}
  \caption{A schematic of the electron acceleration mechanism at the flare loop-top. In a quasi-steady state (top right), the inward plasma flow ($\boldsymbol{u}_{E}$) driven by slow shocks opposes the outward magnetic field gradient ($\nabla B$), resulting in betatron cooling. Conversely, in an unsteady, expanding loop-top (bottom right), the outward plasma flow aligns with the magnetic field gradient, leading to positive betatron acceleration. This difference explains the stronger net energization in the unsteady snapshot analyzed here.}
  \label{fig:acceleration_mechanism}
\end{figure*}

As illustrated in the schematic diagram in Figure~\ref{fig:acceleration_mechanism}, this sign reversal is governed by the direction of the background plasma flow relative to the magnetic field gradient. In the quasi-steady state (top right panel), an inward-directed plasma flow ($\boldsymbol{u}_{E}$) exists, driven by slow shocks. This inward plasma flow and the outward magnetic field gradient ($\nabla B$) are in opposite directions. Consequently, betatron cooling occurs, offsetting the energy gain from Fermi reflection. In the unsteady state (bottom right panel), plasmoid collisions drive an outward-directed plasma flow. This outward plasma flow and the outward magnetic field gradient are in the same direction. Therefore, betatron acceleration occurs.

\section{Summary and Discussion}\label{sec:discussion}

By coupling an MHD simulation with relativistic guiding-center test-particle calculations, we compared electron energization in two loop-top snapshots: a quasi-steady state and an unsteady state perturbed by a plasmoid collision.
Loop-top unsteadiness can be driven by intrinsic oscillations \citep{takasao_above-the-loop-top_2016,wang_current-sheet_2022,Shibata_2023} or by plasmoid collisions \citep{jelinek_oscillations_2017,takasao_observational_2016,ye_role_2020}.
This snapshot comparison focuses on the latter case and indicates that the loop-top state can modulate electron energization. In the quasi-steady state, betatron cooling can offset Fermi reflection gains or even exceed them, producing a net energy loss. By contrast, in the plasmoid-perturbed state, both terms act constructively, producing a net energy gain. The key finding is that betatron energization can reverse its sign when transient changes in the background-flow direction alter how particles sample $\nabla B$.
These results support the importance of time-dependent loop-top dynamics for understanding adiabatic electron energization in solar flares and other explosive magnetic reconnection events.

This modulation is important because the loop-top often represents a critical passage for electrons accelerated in the reconnection region. The compressed magnetic field at the loop-top exit functions as a critical modulation point, energizing or de-energizing electrons en route to the chromosphere. The acceleration or deceleration experienced during this passage influences the electron population that ultimately produces observable HXR and microwave emissions. Our results therefore suggest that models of flare electron transport and energization should take into account the time-dependent plasma dynamics at the loop-top.

Our findings can help to explain several key observational features of loop-top acceleration.
Observations show strong temporal variability in loop-top HXR sources, including quasi-periodic pulsations (QPPs) \citep{mclaughlin_modelling_2018,French_2024, kumar_x-rayradio_2025}. These QPPs are consistent with repeated plasmoid formation and collisions in the reconnection current sheet \citep{liu_plasmoid_2013,takasao_observational_2016,kumar_x-rayradio_2025}. In our model, each plasmoid collision transiently compresses the loop-top field and drives diverging (outward) flows, which in turn produce positive betatron energization. Thus, the observed QPPs in HXR and microwave emissions may directly reflect the modulation of electron acceleration efficiency by these dynamic events.
We note, however, that the effects of plasmoid collisions may vary depending on location and acceleration mechanisms. While our results show enhanced acceleration at the loop-top through betatron energization, the disruption of termination shocks, which could be triggered by plasmoid interactions, has been suggested to reduce shock acceleration efficiency \citep{chen_particle_2015}. This suggests that the net effect of plasmoid collisions on particle acceleration depends on the specific plasma environment and the dominant acceleration process.

Our work also connects to broader frameworks for particle acceleration in reconnecting magnetic fields. Kinetic reconnection studies have separated Fermi reflection and betatron contributions to electron energization, with betatron often producing net cooling when $B$ decreases \citep{Dahlin_2014}. Turbulent reconnecting plasmas can also accelerate particles stochastically \citep{petrosian_stochastic_2012}. Collapsing magnetic trap models provide a closely related context because they describe particle energization by compression and field-line evolution near flare loop-tops \citep{Somov_1997,Birn_2017}. Recent work has further quantified betatron and Fermi contributions in such models \citep{Mowbray_2025}. The present snapshot comparison adds to this context by showing that the betatron contribution can change sign between two MHD loop-top states.

A closely related MHD-plus-test-particle study is \citet{Karlicky_2006}, which followed guiding-center particles in loop-top fields from a quasi-stationary MHD state below a vertical Harris-sheet-type current sheet. The present calculation uses a similar strategy but focuses on a different question. We compare quasi-steady and plasmoid-perturbed MHD snapshots to isolate how the loop-top state changes the local adiabatic energy-change terms. This focus complements their treatment of a quasi-stationary collapsing trap with Coulomb losses, scattering, and X-ray source formation. In contrast, we do not model collisional transport or emission, but show that the betatron term can change sign when a plasmoid collision alters the loop-top flow. The MHD models also differ. Our simulation starts from a force-free Harris current sheet in a gravitationally stratified atmosphere and solves the MHD equations including heat conduction. As a result, the collapsing trap forms self-consistently through the MHD evolution, and its dynamics are further modulated by plasmoid collisions.

Several limitations of the present study should be acknowledged.
First, our test-particle calculations treat the background MHD fields as frozen-in on the particle integration timescale ($t_{\mathrm{GCA}} \sim 1\,\mathrm{s}$), which is less than the Alfv\'en timescale characterizing MHD evolution. Although each snapshot analysis assumes a static background, our snapshot-comparison approach (contrasting the quasi-steady $(t = 130 \mathrm{\ s})$ and unsteady $(t = 138 \mathrm{\ s})$ states) shows that time-dependent loop-top dynamics modulate electron energization. In these fixed snapshots, particle trajectories can appear to cross the magnetic field lines shown in Figures~\ref{fig:steady_combined} and \ref{fig:unsteady_combined}; in fully time-dependent ideal MHD fields, the $\boldsymbol{E} \times \boldsymbol{B}$ drift should advect guiding centers together with evolving field lines to leading order.

Second, our analysis focuses specifically on plasmoid-collision-driven unsteadiness. Other dynamic processes, such as intrinsic loop-top oscillations \citep{takasao_above-the-loop-top_2016, wang_current-sheet_2022, Shibata_2023}, can also modulate the loop-top magnetic field and plasma flows, though quantifying their effects requires dedicated simulations.

Third, our simulations are 2.5-dimensional. While 3D MHD simulations show that loop-top turbulence and oscillations persist \citep{Shibata_2023,shen_origin_2022}, 3D transport effects, such as particle escape along flux ropes, can also significantly alter acceleration efficiency relative to two-dimensional (2D) models \citep{dahlin_role_2017}.

Fourth, our guiding-center test-particle approach omits pitch-angle scattering. Incorporating such scattering would likely enhance high-energy tail formation, as it can trap particles more effectively or enable multiple transits through the acceleration region \citep{kong_modeling_2025}.

Finally, we note both the limitations and complementary advantages of the GCA and MHD approximations. While our approach effectively captures macro-scale acceleration processes such as contracting magnetic traps, it does not include microscopic wave-particle interactions, high-frequency fluctuations, kinetic instabilities, or particle feedback, which require kinetic or multi-scale treatment. The present results should therefore not be interpreted as a fully kinetic model of flare electron acceleration or as a calculation of nonthermal spectral formation. Recent studies using Particle-In-Cell (PIC) simulations have highlighted the importance of such microscopic processes \citep{che_brief_2019, che_electron_2020}. Recent multi-scale modeling efforts have started to bridge the gap between kinetic and fluid scales. For instance, \citet{akutagawa_influence_2025} demonstrated using their multi-hierarchy simulation code that while short-wavelength kinetic waves (e.g., Whistler waves) are confined to the kinetic region, larger-scale MHD structures can smoothly propagate across scales, suggesting that macroscopic dynamics may be largely robust against microscopic fluctuations. On the other hand, \citet{haahr_coupling_2025} developed a PIC solver embedded within an MHD framework to capture non-local kinetic effects in solar flares, highlighting the necessity of self-consistent coupling to resolve the interplay between particle acceleration and macroscopic reconnection dynamics. Further investigations incorporating multi-scale physics will be needed to rigorously verify the impact of microscopic physics on the loop-top acceleration envisioned in our model. In this framework, our study isolates the adiabatic energy gain associated with MHD-scale structural evolution. As demonstrated in preceding studies dealing with complex or turbulent fields \citep[e.g.,][]{Dahlin_2014, gordovskyy_combining_2019, bacchini_particle_2024, oyre_test_2025}, the MHD+GCA approach effectively isolates and quantifies the energy gain driven by macroscopic structural changes, such as the contraction of magnetic traps, which is the specific focus of our study. Our work complements kinetic studies by clarifying the role of macro-scale loop-top dynamics.

%% Acknowledgments
\begin{acknowledgments}
We would like to thank Dr. S. Nagasawa for his helpful comments.
Y.S. acknowledges support from the Junior Fellow Program, run by National Astronomical Observatory of Japan (NAOJ).
This work was supported by JSPS KAKENHI Grant Number JP20K14519 (T.K.), JP21KK0052 and JP22H00134 (N.N.), and JP21H04487 and JP22KK0043 (S.T.).
This work was carried out by the joint research program of Institute for Space-Earth Environmental Research, Nagoya University.
Numerical computations were carried out on Cray XC50 at the Center for Computational Astrophysics, NAOJ.
\end{acknowledgments}

\appendix
\section{Selection Criteria for Example Trajectories} \label{sec:appendix_criteria}

To illustrate the acceleration mechanisms discussed in Sections \ref{subsec:quasi-steady} and \ref{subsec:unsteady}, we selected example trajectories from the $10^7$ test particles using the following criteria.
First, we identified particles whose maximum rate of energy gain from Fermi reflection occurred within the compressed magnetic field at the loop-top edge.
Second, we applied spatial filtering to select particles from both sides of the loop-top.
For the quasi-steady state ($t=130$ s), we used $x < -1$ Mm for left-side particles and $x > +1$ Mm for right-side particles.
For the unsteady state ($t=138$ s), we applied a stricter spatial criterion: $x < -2$ Mm for left-side particles and $x > +2$ Mm for right-side particles.
Third, we retained only particles exhibiting quasi-periodic trapped motion, characterized by a bounce amplitude $\geq 0.1$ Mm.
Fourth, we selected particles showing significant energy gain via Fermi reflection.
For the quasi-steady state, we required Fermi reflection energy gain $\geq 0.5$ keV during the 1-second tracking period.
For the unsteady state, we retained all particles showing positive Fermi reflection energy gain.
Finally, from the particles satisfying all these criteria, we selected the top 1\% by total kinetic energy gain.

This multi-stage filtering progressively reduced the candidate pool. For the quasi-steady state ($t=130$ s), the sequence was: $10^7 \to 3.2 \times 10^5$ (loop-top region) $\to 2.0 \times 10^5$ (spatial filter) $\to 1.9 \times 10^5$ (bounce amplitude) $\to 1.2 \times 10^4$ (Fermi reflection) $\to 118$ (top 1\%). For the unsteady state ($t=138$ s): $10^7 \to 1.6 \times 10^4$ (loop-top region) $\to 1.5 \times 10^4$ (spatial filter) $\to 9.4 \times 10^3$ (bounce amplitude) $\to 2.8 \times 10^3$ (Fermi reflection) $\to 29$ (top 1\%). From each final population, we randomly selected one particle from each side for detailed trajectory analysis.

\section{Population Analysis in the Loop-top-edge Region} \label{sec:appendix_population}

To test whether the energy-decomposition trend inferred from the selected trajectories also appears statistically among particles sampling the loop-top edge, we analyzed a test-particle calculation in which the initial particle positions were restricted to the loop-top region. Each of the quasi-steady and unsteady snapshots contains $4{,}999{,}904$ tracked particles.

\begin{figure*}[!htbp]
  \centering
  \includegraphics[width=0.98\textwidth,height=0.85\textheight,keepaspectratio]{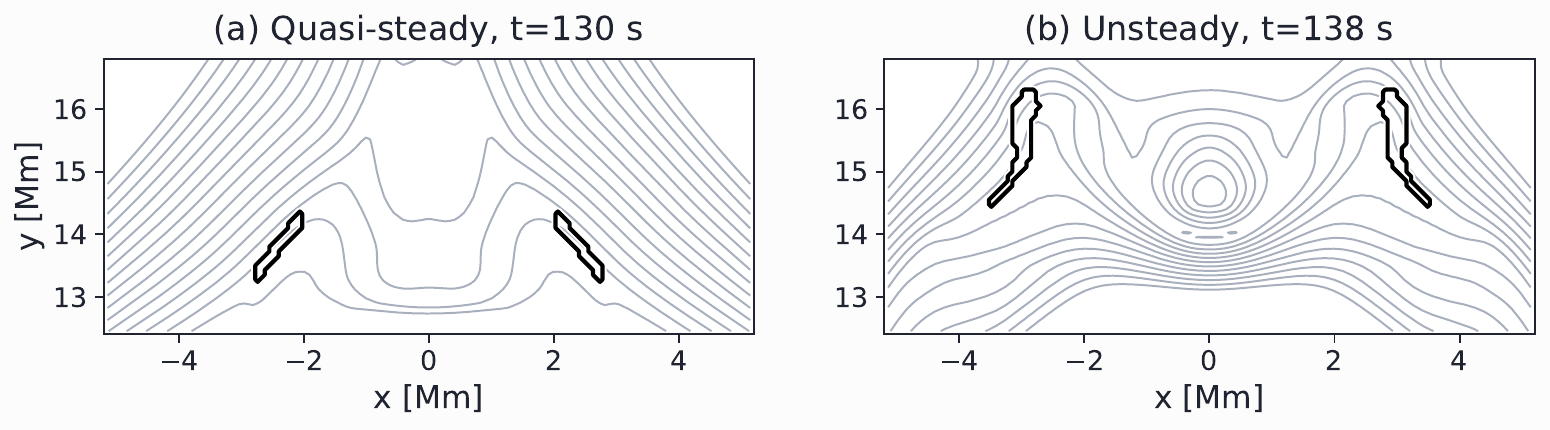}
  \caption{
  Spatial regions used for particle selection in the calculation with loop-top-only initial conditions.
  Panels (a) and (b) show magnetic field lines for the quasi-steady and unsteady snapshots; black contours mark the selected loop-top-edge regions.
  }
  \label{fig:population_active_region}
\end{figure*}

Panels (a) and (b) of Figure~\ref{fig:population_active_region} show the loop-top-edge regions used to select the particle sample. These regions were defined from the MHD fields. We first identified broad loop-top-edge candidate bands from the density and magnetic-field morphology: $2.0 \leq |x| \leq 2.8$ Mm and $13.0 \leq y \leq 14.8$ Mm for the quasi-steady state, and $2.7 \leq |x| \leq 4.2$ Mm and $14.2 \leq y \leq 16.3$ Mm for the unsteady state. Within these bands, we selected loop-top-edge regions using the local relation between the $\boldsymbol{E}\times\boldsymbol{B}$ flow and the magnetic-field gradient along the outward horizontal direction from the loop-top axis, which controls the sign of the betatron term. Specifically, we defined $x_{\mathrm{out}}=|x|$, $u_{E,\mathrm{out}}=\mathrm{sgn}(x)u_{E,x}$, and $\partial B/\partial x_{\mathrm{out}}=\mathrm{sgn}(x)\partial B/\partial x$. For the quasi-steady state, we retained MHD grid cells where the product $(-u_{E,\mathrm{out}})(\partial B/\partial x_{\mathrm{out}})$ was positive, corresponding to inward horizontal flow in a region where $B$ increases outward. For the unsteady state, we retained MHD grid cells where $(+u_{E,\mathrm{out}})(\partial B/\partial x_{\mathrm{out}})$ was positive, corresponding to outward horizontal flow in a region where $B$ increases outward. In each snapshot, we then kept the upper $20\%$ of the positive product values and retained the largest connected component on each side of the loop-top. This procedure defines connected, nonrectangular regions on both sides of the loop-top. Their bounding coordinate ranges are $2.06 \leq |x| \leq 2.74$ Mm and $13.28 \leq y \leq 14.33$ Mm in the quasi-steady state, and $2.74 \leq |x| \leq 3.49$ Mm and $14.48 \leq y \leq 16.28$ Mm in the unsteady state.

After these regions were defined, particles were selected by requiring their sampled trajectories to enter one of the selected loop-top-edge regions and to satisfy the following trajectory criteria. We required an $x$-direction peak-to-peak displacement of at least $0.1$ Mm, at least two turning points in the $x$ direction, and at least $99\%$ of the sampled positions to remain on one side of the loop-top ($x<0$ or $x>0$). No condition was imposed on energy gain.

To compare the selected-particle population with the example trajectories in Table~\ref{tab:acceleration_comparison}, we evaluated the energy change rate per oscillation cycle for all selected particles in each state. The interval between successive local maxima of $|x|$ was treated as one oscillation cycle. A cycle was retained only when it remained on one side of the loop-top, had a measurable $x$-direction excursion, and intersected the selected loop-top-edge region. This yielded $460{,}624$ particles with $754{,}921$ retained cycles in the quasi-steady state and $700{,}993$ particles with $1{,}738{,}924$ retained cycles in the unsteady state. For each particle, the retained cycle changes were first summed and then normalized by the summed mean kinetic energy over the same cycles,
\begin{equation}
R_{X,p}=100\,
\frac{\sum_c \Delta E_{X,p,c}}
{\sum_c \langle E\rangle_{p,c}},
\label{eq:population_cycle_rate}
\end{equation}
Here, $p$ labels an individual particle, $c$ labels one retained oscillation cycle of that particle, and $X$ denotes the net kinetic, Fermi-reflection, or betatron contribution. The quantity $\Delta E_{X,p,c}$ is the energy change in contribution $X$ over cycle $c$, and $\langle E\rangle_{p,c}$ is the cycle-averaged total kinetic energy. This common total-kinetic-energy normalization places the three terms on the same scale for this population comparison and avoids overweighting cycles with small parallel or perpendicular energies.

\begin{deluxetable}{lccc}[!htbp]
  \tablecaption{Cycle-based energy change rates for selected particles in the loop-top-edge region
  \label{tab:population_cycle_rates}}
  \tablehead{
      \colhead{State} & \multicolumn{3}{c}{Rate per cycle (\% of kinetic energy)} \\
      \cline{2-4}
      & \colhead{Net kinetic} & \colhead{Fermi reflection} & \colhead{Betatron}
  }
  \startdata
      Quasi-steady ($t=130\,\mathrm{s}$) & $-6.2\,[-12.5,-2.3]$ & $+9.3\,[-0.3,+15.7]$ & $-15.5\,[-26.1,-7.4]$ \\
      Unsteady ($t=138\,\mathrm{s}$) & $+6.9\,[-1.3,+17.6]$ & $+1.9\,[-0.8,+5.8]$ & $+5.6\,[-5.6,+16.2]$ \\
  \enddata
  \tablecomments{Rates were calculated for each particle after summing over its retained cycles using Equation~\ref{eq:population_cycle_rate}. For each column, values are medians across particles; bracketed values indicate the 5th--95th percentile ranges.}
\end{deluxetable}

The resulting rates in Table~\ref{tab:population_cycle_rates} are consistent with the energy-decomposition trend found for the selected example trajectories in Table~\ref{tab:acceleration_comparison}. In the quasi-steady state, the Fermi reflection contribution is positive, but the betatron contribution is negative and larger in magnitude, yielding a negative net rate. In the unsteady state, the betatron contribution changes sign and the net rate becomes positive.

\clearpage

\bibliography{main}
\bibliographystyle{aasjournalv7}

\end{document}